\input harvmac
\input epsf
\rightline{RI-05-02}
\Title{
\rightline{hep-th/0212242}
}
{\vbox{\centerline{Removing Singularities}}}
\medskip

\centerline{\it Shmuel Elitzur, Amit Giveon, Eliezer Rabinovici}
\bigskip
\centerline{Racah Institute of Physics, The Hebrew University,
Jerusalem, 91904, Israel}

\smallskip

\vglue .3cm
\bigskip

\noindent
Big bang/crunch curvature 
singularities in exact CFT string backgrounds can be 
removed by turning on gauge fields. This is described within a family of 
${SL(2)\times SU(2)\times U(1)_x\over U(1)\times U(1)}$ quotient CFTs.
Uncharged incoming wavefunctions from the ``whiskers'' 
of the extended universe
can be fully reflected if and only if a big bang/crunch curvature singularity, 
from which they are scattered, exists. 
Extended BTZ-like singularities remain as long as $U(1)_x$ is compact.

\Date{10/02}

\def\journal#1&#2(#3){\unskip, \sl #1\ \bf #2 \rm(19#3) }
\def\andjournal#1&#2(#3){\sl #1~\bf #2 \rm (19#3) }

\def\frac#1#2{{#1\over#2}}

\def\inbar{\,\vrule height1.5ex width.4pt depth0pt}
\def\IC{\relax\hbox{$\inbar\kern-.3em{\rm C}$}}
\def\IR{\relax{\rm I\kern-.18em R}}
\def\IP{\relax{\rm I\kern-.18em P}}

%
%

%
\catcode`\@=11
\def\slash#1{\mathord{\mathpalette\c@ncel{#1}}}
\overfullrule=0pt

\def\underrel#1\over#2{\mathrel{\mathop{\kern\z@#1}\limits_{#2}}}

\catcode`\@=12


%

\def \sinh{{\rm sinh}}
\def \cosh{{\rm cosh}}


\def\s{{\bf S}}

\newsec{Introduction}

\lref\gpr{
A.~Giveon, M.~Porrati and E.~Rabinovici,
``Target space duality in string theory,''
Phys.\ Rept.\  {\bf 244}, 77 (1994)
[arXiv:hep-th/9401139].
}

\lref\nw{
C.~R.~Nappi and E.~Witten,
``A Closed, expanding universe in string theory,''
Phys.\ Lett.\ B {\bf 293}, 309 (1992)
[arXiv:hep-th/9206078].
}
\lref\egkr{
S.~Elitzur, A.~Giveon, D.~Kutasov and E.~Rabinovici,
``From big bang to big crunch and beyond,''
JHEP {\bf 0206}, 017 (2002)
[arXiv:hep-th/0204189].
}
\lref\BarsDX{
I.~Bars and K.~Sfetsos,
``$SL(2,\IR)\times SU(2)/\IR^2$ 
string model in curved space-time and exact conformal results,''
Phys.\ Lett.\ B {\bf 301}, 183 (1993)
[arXiv:hep-th/9208001].
}
\lref\gp{
A.~Giveon and A.~Pasquinucci,
``On cosmological string backgrounds with toroidal isometries,''
Phys.\ Lett.\ B {\bf 294}, 162 (1992)
[arXiv:hep-th/9208076].
}
\lref\gmv{
M.~Gasperini, J.~Maharana and G.~Veneziano,
``Boosting away singularities from conformal string backgrounds,''
Phys.\ Lett.\ B {\bf 296}, 51 (1992)
[arXiv:hep-th/9209052].
}
\lref\vil{
N.J. Vilenkin, ``Special Functions and the Theory of Group Representations''
AMS, 1968.
}
\lref\btz{
M.~Banados, C.~Teitelboim and J.~Zanelli,
``The Black Hole In Three-Dimensional Space-Time,''
Phys.\ Rev.\ Lett.\  {\bf 69}, 1849 (1992)
[arXiv:hep-th/9204099].
}
\lref\bhtz{
M.~Banados, M.~Henneaux, C.~Teitelboim and J.~Zanelli,
``Geometry of the (2+1) black hole,''
Phys.\ Rev.\ D {\bf 48}, 1506 (1993)
[arXiv:gr-qc/9302012].
}
\lref\mm{
E.~J.~Martinec and W.~McElgin,
``Exciting AdS orbifolds,''
arXiv:hep-th/0206175.
}
\lref\fs{
J.~Figueroa-O'Farrill and J.~Simon,
``Generalized supersymmetric fluxbranes,''
JHEP {\bf 0112}, 011 (2001)
[arXiv:hep-th/0110170].
}
\lref\s{
J.~Simon,
``The geometry of null rotation identifications,''
JHEP {\bf 0206}, 001 (2002)
[arXiv:hep-th/0203201].
}
\lref\HorowitzSR{
G.~T.~Horowitz and A.~R.~Steif,
``Strings In Strong Gravitational Fields,''
Phys.\ Rev.\ D {\bf 42}, 1950 (1990).
}
\lref\HorowitzAP{
G.~T.~Horowitz and A.~R.~Steif,
``Singular String Solutions With Nonsingular Initial Data,''
Phys.\ Lett.\ B {\bf 258}, 91 (1991).
}
\lref\KhouryBZ{
J.~Khoury, B.~A.~Ovrut, N.~Seiberg, P.~J.~Steinhardt and N.~Turok,
``From big crunch to big bang,''
Phys.\ Rev.\ D {\bf 65}, 086007 (2002)
[arXiv:hep-th/0108187].
}
\lref\SeibergHR{
N.~Seiberg,
``From big crunch to big bang - is it possible?,''
arXiv:hep-th/0201039.
}
\lref\BalasubramanianRY{
V.~Balasubramanian, S.~F.~Hassan, E.~Keski-Vakkuri and A.~Naqvi,
``A space-time orbifold: A toy model for a cosmological singularity,''
arXiv:hep-th/0202187.
}
\lref\CornalbaFI{
L.~Cornalba and M.~S.~Costa,
``A New Cosmological Scenario in String Theory,''
Phys.\ Rev.\ D {\bf 66}, 066001 (2002)
[arXiv:hep-th/0203031].
}
\lref\NekrasovKF{
N.~A.~Nekrasov,
``Milne universe, tachyons, and quantum group,''
arXiv:hep-th/0203112.
}
\lref\KiritsisKZ{
E.~Kiritsis and B.~Pioline,
``Strings in homogeneous gravitational waves and null holography,''
JHEP {\bf 0208}, 048 (2002)
[arXiv:hep-th/0204004].
}
\lref\TolleyCV{
A.~J.~Tolley and N.~Turok,
``Quantum fields in a big crunch / big bang spacetime,''
arXiv:hep-th/0204091.
}
\lref\LiuFT{
H.~Liu, G.~Moore and N.~Seiberg,
``Strings in a time-dependent orbifold,''
JHEP {\bf 0206}, 045 (2002)
[arXiv:hep-th/0204168].
}
\lref\CornalbaNV{
L.~Cornalba, M.~S.~Costa and C.~Kounnas,
``A resolution of the cosmological singularity with orientifolds,''
Nucl.\ Phys.\ B {\bf 637}, 378 (2002)
[arXiv:hep-th/0204261].
}
\lref\LawrenceAJ{
A.~Lawrence,
``On the instability of 3D null singularities,''
arXiv:hep-th/0205288.
}
\lref\LiuKB{
H.~Liu, G.~Moore and N.~Seiberg,
``Strings in time-dependent orbifolds,''
JHEP {\bf 0210}, 031 (2002)
[arXiv:hep-th/0206182].
}
\lref\HorowitzMW{
G.~T.~Horowitz and J.~Polchinski,
``Instability of spacelike and null orbifold singularities,''
arXiv:hep-th/0206228.
}
\lref\HoravaAM{
P.~Horava,
``Some exact solutions of string theory 
in four-dimensions and five-dimensions,''
Phys.\ Lett.\ B {\bf 278}, 101 (1992)
[arXiv:hep-th/9110067].
}
\lref\KounnasWC{
C.~Kounnas and D.~Lust,
``Cosmological string backgrounds from gauged WZW models,''
Phys.\ Lett.\ B {\bf 289}, 56 (1992)
[arXiv:hep-th/9205046].
}
\lref\CrapsII{
B.~Craps, D.~Kutasov and G.~Rajesh,
``String propagation in the presence of cosmological singularities,''
JHEP {\bf 0206}, 053 (2002)
[arXiv:hep-th/0205101].
}
\lref\TroostWK{
J.~Troost,
``Winding strings and AdS(3) black holes,''
JHEP {\bf 0209}, 041 (2002)
[arXiv:hep-th/0206118].
}
\lref\FabingerKR{
M.~Fabinger and J.~McGreevy,
``On smooth time-dependent orbifolds and null singularities,''
arXiv:hep-th/0206196.
}
\lref\BuchelKJ{
A.~Buchel, P.~Langfelder and J.~Walcher,
``On time-dependent backgrounds in supergravity and string theory,''
arXiv:hep-th/0207214.
}
\lref\HemmingKD{
S.~Hemming, E.~Keski-Vakkuri and P.~Kraus,
``Strings in the extended BTZ spacetime,''
JHEP {\bf 0210}, 006 (2002)
[arXiv:hep-th/0208003].
}
\lref\SimonCF{
J.~Simon,
``Null orbifolds in AdS, time dependence and holography,''
arXiv:hep-th/0208165.
}
\lref\GasperiniBN{
M.~Gasperini and G.~Veneziano,
``The pre-big bang scenario in string cosmology,''
arXiv:hep-th/0207130.
}
\lref\brs{
K.~Bardakci, E.~Rabinovici and B.~Saering,
``String Models With $C<1$ Components,''
Nucl.\ Phys.\ B {\bf 299}, 151 (1988).
}
\lref\abr{
D.~Altschuler, K.~Bardakci and E.~Rabinovici,     
``A Construction Of The $C<1$ Modular Invariant Partition Functions,''
Commun.\ Math.\ Phys.\  {\bf 118}, 241 (1988).
}
\lref\n{
W.~Nahm,  
``Gauging Symmetries Of Two-Dimensional Conformally Invariant Models,''
UCD-88-02. 
}
\lref\gku{
K.~Gawedzki and A.~Kupiainen,
``Coset Construction From Functional Integrals,''
Nucl.\ Phys.\ B {\bf 320}, 625 (1989). 
}   
\lref\kpsy{
D.~Karabali, Q.~H.~Park, H.~J.~Schnitzer and Z.~Yang,
``A Gko Construction Based On A Path Integral Formulation Of Gauged
Wess-Zumino-Witten Actions,''
Phys.\ Lett.\ B {\bf 216}, 307 (1989).   
}
\lref\dvv{
R.~Dijkgraaf, H.~Verlinde and E.~Verlinde,
``String propagation in a black hole geometry,''
Nucl.\ Phys.\ B {\bf 371}, 269 (1992).
}
\lref\HiscockVQ{         
W.~A.~Hiscock and D.~A.~Konkowski,    
``Quantum Vacuum Energy In Taub - Nut (Newman-Unti-Tamburino) Type
Cosmologies,''  
Phys.\ Rev.\ D {\bf 26}, 1225 (1982).
}

Studies of the propagation of strings in time-dependent backgrounds have
highlighted several basic issues. One is related to the possibility that
extended objects such as strings may propagate beyond space-like
singularities. Another concerns the manner one may embed a 
compact cosmology in an allowed perturbative string background. A third
related issue involves the appropriate boundary conditions to be imposed
in the presence of such singularities. A way to evaluate an entropy in
string cosmology was addressed as well.    
These difficult issues were investigated in exact perturbative CFT
backgrounds.
The first issue was addressed in two types of CFT backgrounds. 
One set of string backgrounds consists of orbifolds of $\IR^{1,d-1}$
\refs{\HorowitzSR,\HorowitzAP,\KhouryBZ,\fs,\SeibergHR,\BalasubramanianRY,
\CornalbaFI,\NekrasovKF,\s,\KiritsisKZ,\TolleyCV,\LiuFT,\CornalbaNV,
\LawrenceAJ,\LiuKB,\FabingerKR,\HorowitzMW,\SimonCF}.
Such space-times are flat away from the singularities.
In the second class of CFT backgrounds 
\refs{\HoravaAM,\KounnasWC,\nw,\BarsDX,\gp,\gmv,\egkr,\CrapsII,\BuchelKJ}, 
gauged WZW models \refs{\brs,\abr,\n,\gku,\kpsy} which involve
$AdS_3\sim SL(2,\IR)$, there are two types of singularities. The
first is also of an orbifold nature, 
in these cases either an $\IR^{1,1}/$Boost or an extended BTZ-like 
singularity \refs{\btz,\bhtz,\TroostWK,\mm,\HemmingKD}, 
whose global structure is that of AdS. 
The second is a Ricci curvature singularity where also the dilaton diverges. 
In what follows we focus on the quotient CFT background investigated in \egkr.

The attitude taken in these studies was to assume the validity of string
perturbation theory, and to compute physical quantities by standard
worldsheet techniques.
Several intriguing results emerged \egkr. 
Strings were found to propagate through these singularities. 
Each expanding universe is connected to a pre collapsing universe at their
corresponding big bang/crunch singularities -- a pre big bang scenario
(for a review, see \GasperiniBN). 
Moreover, the compact cosmological models were found to be accompanied by 
static ``whiskers,'' which have a non-compact direction with a
space-like linear dilaton. 
In the whiskers an $S$-matrix setup is possible.
Boundary conditions can be set in a weak coupling asymptotic regime, 
determining the boundary conditions at the singularities. 
The entropy was found to be significantly reduced 
(relative to other theories with global AdS structure 
or with an asymptotic space-like linear dilaton). 
The whiskers also include closed time-like curves
and time-like domain wall singularities.

In this work we present a larger class of models in which
compact cosmologies are once again embedded in a non-compact 
space-time which includes whiskers.
In this family of models one can gradually 
separate the curvature singularities 
from the BTZ-like singularities. The big bang/crunch curvature singularity
can be removed by pushing the domain wall connected to it 
towards the boundary of the whisker, leaving behind an extended 
BTZ-like singularity at the times when a compact universe meets a whisker.
The nature of the singularity may play a role in the important question
regarding the validity of perturbation theory.
A general argument is that as the universe collapses, a large amount of
energy will eventually be concentrated near the space-like singularity,
causing a large back reaction on the structure of the singularity
\refs{\HiscockVQ,\LawrenceAJ,\HorowitzMW}. 
As a consequence, this may affect the validity of perturbation theory
\refs{\LiuKB,\HorowitzMW}. 
The extended BTZ-like singularities which appear in our family of quotient
CFTs have an AdS structure which might affect the strength of
the back reaction, and as a consequence, the question
regarding the validity of perturbation theory.

The time-dependent string background studied in \egkr\ is based on the  
${SL(2,\IR)\times SU(2)\over U(1)\times U(1)}$ quotient CFT \nw.
As mentioned, it consists of a sequence of closed, expanding and recollapsing
universes, each connected at its big bang and big crunch singularities 
to the whiskers \egkr.
Observables in this string background are of two kinds.
Wavefunctions localized near the closed universes correspond to vertex
operators in the quotient CFT which are exponentially supported at the boundary
of the whiskers. On the other hand, scattering states correspond to
delta-function normalizable vertex operators.
{}From the latter, one can construct linear combinations
which describe incoming waves prepared in a certain whisker
and scattered from the big bang/crunch singularities.
Generically, these waves are partially reflected from the singularity. 
However, it was found that one can always prepare scattering waves which
are fully reflected from the singularity. These are regular combinations
of physical wavefunctions each of which develops a logarithmic singularity
at the corresponding big bang/crunch.  

In this note, we show that by turning on an Abelian gauge field 
one can remove the curvature singularities in the time-dependent background of 
\refs{\nw,\egkr}. Removing such singularities was done previously by
$O(d,d,\IR)$ rotations of 
${SL(2,\IR)\times SU(2)\over U(1)\times U(1)}\times U(1)_x$
in \refs{\gp,\gmv} (for a review, see \gpr). 
By turning on gauge fields one can remove the big bang and/or the big crunch 
curvature singularities.
We show that by gradually turning on a gauge field one can
``push'' a big bang/crunch singularity and the domain wall singularity to 
which it is connected in the whisker (see \egkr\ for details)
towards the boundary of space-time. 
For particular values of the gauge field the background 
has no curvature singularity. On the other hand, extended BTZ-like orbifold
singularities remain, unless the extra fifth dimension $U(1)_x$ 
is non-compact.

Here we study the cosmology of \nw\
in the presence of a gauge field by considering
a three-parameter subfamily of the 
${SL(2,\IR)\times SU(2)\times U(1)\over U(1)\times U(1)}$ quotient CFTs.
This allows us to consider physical vertex operators in such backgrounds.
We find that fully reflected uncharged wavefunctions exist if and only if
there is a big bang or big crunch
singularity from which they are scattered. 

In section 2, we describe the geometry of the 
${SL(2,\IR)\times SU(2)\times U(1)\over U(1)\times U(1)}$
sigma-model and the wavefunctions in this CFT background.
In section 3, we discuss the singularities, 
their removal, and the corresponding
behavior of the wavefunctions. Our main results are summarized in section 4.
In appendix A, we present the relation between
singularities in quotient CFT backgrounds and fixed points
in the underlying sigma-model under a subgroup 
of the gauged isometry group.

\newsec{$[SL(2,\IR)\times SU(2)\times U(1)]/U(1)^2$}

\subsec{Geometry}

We construct a $5$-dimensional time-dependent background
by gauging the WZW model of the $7$-dimensional 
$SL(2,\IR)\times SU(2)\times U(1)$ group manifold 
by a non-compact space-like $U(1)\times U(1)$ subgroup. 
Let $(g,g',x)$ be a point on the product group manifold and let $k$
and $k'$ be the levels of $SL(2,\IR)$ and $SU(2)$ respectively.
Here $g \in SL(2,\IR)$, $g'\in SU(2)$ and $x$ denotes a point 
on the unit circle. The $U(1)^2$ gauge group acts as
\eqn\gatr{(g,g',x_L,x_R)\rightarrow (e^{\rho \sigma_{3}/\sqrt{k} }g
 e^{\tau \sigma_{3}/\sqrt{k}}
, e^{i\rho' \sigma_{3}/\sqrt{k'} }g' e^{i\tau' \sigma_{3}/\sqrt{k'}},
x_L+\rho'',x_R+\tau'')~.}
Since we gauge only $U(1)^2$ out of the three $U(1)$ right-handed
generators  in \gatr, the three parameters $\tau,\tau'$ and $\tau''$
are not independent but rather are constrained by
\eqn\constr{\underline v\cdot  \underline \tau 
\equiv v_1\tau + v_2 \tau' + v_3\tau'' = 0~,}
where $\underline v$ is some real $3$-vector.
The left-handed parameters $\rho,\rho'$ and $\rho''$ in \gatr\ 
depend linearly on the right-handed $\tau$ parameters. For an anomaly 
free gauging this dependence has to take the form 
\eqn\orth{ \underline{ \rho} = R \underline {\tau }~,}
where $\underline \rho$ is a $3$-vector with $\rho ,\rho'$ and $\rho''$ as
components, similarly for $\underline \tau$,
and $R$ is a $3 \times 3$ orthogonal
matrix. Apart from $k$, $k'$ and the radius of the circle 
parametrized by $x$, our model depends then on $5$ parameters,
three of them fixing the matrix $R$ in \orth, and two more fixing the
vector $\underline v$,  which is defined by \constr\ 
only up to a multiplication by a scalar.

To perform the gauging \gatr\  one introduces dynamical fields 
$\hat{ \underline{\rho}},
\hat{\underline {\tau}} $ corresponding to the parameters $\underline{\rho},
\underline {\tau} $, subject to the constraints
\eqn\dconstr{\hat{\underline{ \tau}} \cdot \underline v 
=\hat{ \underline{\rho}} \cdot R \underline v =0~.}
The gauged action is then defined by 
\eqn\gact{\eqalign{S=&S[e^{\hat\rho \sigma_{3}/\sqrt{k} }g
 e^{\hat\tau \sigma_{3}/\sqrt{k}}]+
S'[ e^{i\hat\rho' \sigma_{3}/\sqrt{k'} }g' 
e^{i\hat\tau' \sigma_{3}/\sqrt{k'}}]+
S''[x+\hat \rho''+\hat \tau'']\cr&-
{1\over {2\pi}}\int d^2z (\partial\hat{ \underline {\rho}}
-R\partial\hat {\underline{ \tau}})^T
(\bar{\partial} \hat{\underline {\rho}}
 -R\bar {\partial} \hat{\underline {\tau}})~.}}
 $S[g]$ is the WZW action of $g\in SL(2)$,
\eqn\wzg{S[g]={k \over {4 \pi}}[\int_{\Sigma} Tr(g^{-1}\partial g
g^{-1}\bar \partial g)-{1 \over 3} \int_{B} Tr (g^{-1}dg)^3 ]~,}
where $\Sigma$ is the string's worldsheet and $B$ a $3$-submanifold 
of  the group $SL(2)$ bounded by the image of $\Sigma$.
 $S'[g']$ is similarly defined for the group $SU(2)$, 
\eqn\wzgp{S'[g']=-{k' \over {4 \pi}}[\int_{\Sigma} Tr(g'^{-1}\partial g'
g'^{-1}\bar \partial g')-{1 \over 3} \int_{B} Tr (g'^{-1}dg')^3 ]~.}
Finally 
\eqn\uone{S''[x]={1 \over {2 \pi}}\int_{\Sigma}\partial x \bar \partial x~.}
Apart from the constraints \dconstr,
$\hat{ \underline{\rho}}$ and $ \hat{\underline{ \tau}}$
are independent fields.
The action \gact\ is invariant under the gauge transformation \gatr\
for the fields $g, g'$ and $x$ together with the transformation
\eqn\agatr{\eqalign{&\hat{ \underline{\rho}}\rightarrow
\hat {\underline{\rho}}-\underline {\rho}
\cr&\hat{\underline{ \tau}}\rightarrow
\hat {\underline{\tau}}-\underline{\tau}}}
provided that the parameters $\underline{\rho}$ and $\underline{\tau}$ satisfy 
the relation \orth. Using the Polyakov-Wiegmann identity one sees that
the action \gact\ depends on $\hat{\underline{\rho}}$
and $\hat{\underline{\tau}}$ only through the quantities
\eqn\af{\eqalign{&{\bf A}=\partial\hat{\underline{\tau}}\cr
&\bar{\bf A}=\bar{\partial}\hat{\underline{\rho}}}}   
The gauged action has then the form
\eqn\act{S= S[g]+S'[g']+S''[x]+{1 \over {2\pi}}\int d^2z[ \bar{\bf J}^T {\bf A}
+\bar{\lambda} {\bf v}^T {\bf A}
+\bar{{\bf A}}^T {\bf J}+\lambda\bar{{\bf A}}^T R{\bf v}
+2\bar{\bf A}^T M{\bf A}]}
Here,
${\bf A}^T$ is the row vector $(A, A', A'')$ defined in \af\ 
consisting of the 
holomorphic components of the gauge fields of $SL(2), SU(2)$ and $U(1)$,
respectively, with a similar definition for $\bar {\bf A}$.
${\bf J}^T$ is the row vector of the currents,
\eqn\cur{{\bf J}^T= (\sqrt{k}Tr[ \partial g g^{-1}\sigma_3],
 -i\sqrt{k'}Tr[ \partial g'g'^{-1}\sigma_3], 2\partial x)}
Similarly, $\bar {\bf J}^T$ is
\eqn\curb{\bar {\bf J}^T= (\sqrt{k} Tr[g^{-1} \bar{\partial} g\sigma_3], 
-i\sqrt{k'}Tr[g'^{-1}\bar{ \partial} g'\sigma_3],2\bar{ \partial} x)}
The $3\times 3$ matrix $M$ in \act\ is of the form,
\eqn\qfo{M=\left(\matrix{{1\over 2}Tr[g^{-1}\sigma_3 g\sigma_3]
&0&0\cr 0&{1\over 2}Tr[g'^{-1}\sigma_3 g'\sigma_3]
&0\cr 0&0&1\cr}\right) + R }
$\lambda$ and $\bar{\lambda}$ are Lagrange multiplyers enforcing 
the constraint corresponding to \dconstr\ on $\bar{\bf A}$ and ${\bf A}$.

Since \act\ is invariant under the gauge transformation \gatr\ and \agatr,
 integrating out the fields ${\bf A},\bar {\bf A}$ leaves an action 
depending on $g,g',x$, invariant under \gatr. Fixing this gauge invariance
results in a $5$-dimensional sigma-model action containing the geometrical
information of the resulting space-time. 

Parametrize $g$ as 
\eqn\para{g=e^{\alpha \sigma_3}g(\theta_i)e^{\beta \sigma_3}}
This parametrization is valid for any matrix $g$ with non-zero elements.
The definition of the factor $g(\theta_i)$ depends on the region 
where $g$ is in the $SL(2)$ group manifold \vil,\egkr. Defining
\eqn\w{W=Tr(\sigma_3 g \sigma_3 g^{-1})}
 $g(\theta_1)$ stands for $e^{i\theta_1 \sigma_2}$ in regions of $SL(2)$
for which $W$ satisfies $|W|\le 2$.
The points of  $SL(2)$ for which $W>2$ are divided into $4$ regions. There 
the factor $g(\theta_2)$ represents
 $\pm e^{\pm\theta_2 \sigma_1}$. For the $4$ regions where 
$W<-2$, $g(\theta_3)=\pm i\sigma_2  e^{\pm\theta_3 \sigma_1}$.
At the point $\theta_1=0$, $W=2$. Here two of the regions 
parametrized by $\theta_2$ meet the region parametrized by $\theta_1$.
Similarly, at $\theta_1=\pi$ the other two regions parametrized by
 $\theta_2$ meet the region parametrized by $\theta_1$.
At $\theta_1={\pi \over 2}$ $(W=-2)$ two regions parametrized by $\theta_3$
meet the $\theta_1$ region and at $\theta_1={3\pi \over 2}$
the other two  $\theta_3$ regions meet  the $\theta_1$ region.
The range of $\theta_{2,3}$ is $0\le \theta_{2,3} < \infty$.
For the group $SL(2)$, $\theta_1$ satisfies $0 \le \theta_1 \le 2\pi$.
For the infinite cover of $SL(2)$, $\theta_1$ satisfies
$-\infty < \theta_1 < \infty$.  
   
Parametrize $g'$ by the Euler angles 
\eqn\parap{g'=e^{i\alpha' \sigma_3}e^{i\theta' \sigma_2}
e^{i\beta' \sigma_3}}
with $0 \le \alpha'<2 \pi, 0 \le \theta' \le {\pi \over 2}, 0 \le \beta'< \pi$.
In these terms the currents in \cur\ and \curb, in regions
where $|W|\le 2$, take the form
\eqn\cocur{{\bf J}^T= \left(2\sqrt{k}(\partial \alpha + 
cos(2\theta_1)\partial \beta),
2 \sqrt{k'}(\partial \alpha' + 
cos(2\theta')\partial \beta'), 2\partial x\right)}
\eqn\cocurb{\bar {\bf J}^T= \left(2\sqrt{k} (\bar {\partial} \beta
+ cos(2\theta_1)\bar {\partial}\alpha), 
2\sqrt{k'}(\bar {\partial} \beta'
+ cos(2\theta')\bar {\partial}\alpha'),2\bar{ \partial} x\right)}
In the regions where $W>2$, $\theta_1$ in \cocur, \cocurb\ 
should be replaced by $i\theta_{2}$. In the regions with $W<-2$,
substitute $i\theta_3$ for $\theta_1 - {\pi\over 2}$.

Fix the $U(1)\times U(1)$ gauge by the condition,
\eqn\gfix{\alpha=\beta=0~,}
which is possible wherever the parametrization \para\ is valid.
The resulting $5$-dimensional manifold is parametrized by 
$(\theta_i, \theta',\alpha',\beta',x)$. Generically, at points where
the parametrization \para\ is valid the manifold looks like a family
of $SU(2)\times S^1$ manifolds parametrized by the continuous
parameter $\theta_i$ in each of the regions of $SL(2)$ described above.
At points corresponding to $g \in SL(2)$ with vanishing elements, including
the points $\theta_1=0,{\pi \over 2},\pi,{3\pi \over 2}$ where various
regions meet, the gauge fixing \gfix\ is not complete and has 
to be supplemented by a gauge condition on the $SU(2)$ or $U(1)$ part.
Above these points the  $SU(2)\times S^1$ manifold will be 
twisted by some extra identification.    

The fixing \gfix\ sets the first components of ${\bf J}$ and $\bar{\bf J}$ 
in \cocur\ and \cocurb\ to zero.
Substituting \gfix\ into \act\ then integrating out ${\bf A},{\bf \bar A},
\lambda$ and $\bar \lambda$ one gets the sigma-model action for
the $5$-dimensional target space as,
\eqn\siact{\eqalign{S&=\int d^2z\Big\{-{k\over{2\pi}}
\partial \theta_1 \bar {\partial} \theta_1 
+{k'\over{2\pi}}\left(\partial \theta' \bar {\partial}\theta'  
+\partial\alpha' \bar {\partial}\alpha' 
+ \partial \beta' \bar {\partial}\beta' 
+ 2cos(2\theta')\partial \alpha' \bar {\partial} \beta'\right)
+{1 \over {2 \pi}}\partial x \bar {\partial}x\cr
-&{k'\over{\pi}} \Big[(M^{-1})_{2,2} - {1\over{v_m( M^{-1}R)_{m,n}v_n}}
 (M^{-1}R)_{2,k}
v_kv_l M^{-1}_{l,2}\Big] (\partial \alpha' + cos(2\theta') \partial \beta')
(\bar {\partial} \beta' + cos(2\theta')\bar {\partial} \alpha')\cr-&
{\sqrt{k'}\over{\pi}}\Big[(M^{-1})_{2,3}- {1\over{v_m (M^{-1}R)_{m,n}v_n}}
( M^{-1}R)_{2,k}
v_kv_l M^{-1}_{l,3}\Big]\partial x(\bar{\partial}\beta' + cos(2\theta')\bar
{\partial}\alpha')\cr
-&{\sqrt{k'}\over{ \pi}}\Big[(M^{-1})_{3,2}- {1\over{v_m (M^{-1}R)_{m,n}v_n}}
 (M^{-1}R)_{3,k}
v_kv_l M^{-1}_{l,2}\Big](\partial \alpha' +
cos(2\theta') \partial \beta')\bar {\partial} x\cr -&
{1\over{\pi}}\Big[(M^{-1})_{3,3}
- {1\over{v_m (M^{-1}R)_{m,n}v_n}}( M^{-1}R)_{3,k}
v_kv_l M^{-1}_{l,3}\Big]\partial x \bar {\partial} x\Big\}}}
The dilaton field $\Phi$, defined such that the string coupling $g_s$ equals
$e^\Phi$, becomes in this geometry
\eqn\dil{\Phi = \Phi_0-{1\over 2}
\left[log( detM)+log(v_m (M^{-1}R)_{m,n}v_n)\right]}
In regions where $W>2$, $\theta_1$ in \siact\ 
should be replaced by $i\theta_{2}$. Similarly,
in regions where $W<-2$, ${\pi\over 2}-\theta_1$  
should be replaced by $i\theta_3$.

We focus on geometries which are essentially $4$-dimensional, namely,
those for which the length of the extra circle parametrized by $x$
is constant. This  will emerge if the parameters of the model
are chosen such that the last term in \siact, proportional to
$\partial x \bar {\partial} x$, vanishes. This happens when we chose 
the vector $\underline v$ of the form
\eqn\hetv{\underline{ v}^T=(0,0,1)}
This means, by \constr, that no gauging is applied to $x_R$.
 Note that such a gauging can also be applied to heterotic backgrounds
where the $U(1)$ symmetry may exist only from the left with no 
right-handed part from the start. Note also that with the choice \hetv, 
in addition to the term proportional to  $\partial x \bar {\partial} x$,
also the term proportional to 
$(\partial\alpha' + cos(2\theta')\partial\beta')\bar {\partial} x$ 
vanishes. The condition \hetv\ selects
a $3$-parameter family of models parametrized by $R$.

The dilaton field of \dil\ for this case is
\eqn\hdilr{\Phi = \Phi_0-{1\over 2}log \left[1
+R_{3,3} cos(2\theta_1)cos(2\theta') +R_{1,1}cos(2\theta_1)
+R_{2,2}cos(2\theta')\right]}
The action \siact\ becomes 
\eqn\hactr{\eqalign{S&=\int d^2z\Big\{-{k\over{2\pi}}
\partial \theta_1 \bar {\partial} \theta_1 
+{k'\over{2\pi}}\left(\partial \theta' \bar {\partial}\theta'  
+\partial\alpha' \bar {\partial}\alpha' 
+ \partial \beta' \bar {\partial}\beta' 
+ 2cos(2\theta')\partial \alpha' \bar {\partial} \beta'\right)
+{1 \over {2 \pi}}\partial x \bar {\partial}x\cr
-&{k'\over{\pi}}\Big[{1 \over (cos(2\theta_1)+cos(2\theta'))
(R_{1,1}+R_{2,2}) + (1+cos(2\theta_1)cos(2\theta'))(1+R_{3,3})}\cr& 
\times \bigl[R_{1,1}+R_{2,2}+cos(2\theta_1)(1+R_{3,3})
-e^{2(\Phi-\Phi_0)}(cos(2\theta_1)R_{2,3}-R_{3,2})
(R_{2,3}-cos(2\theta_1)R_{3,2})\bigr]\Big]\cr&\times (\partial \alpha' +
 cos(2\theta') \partial \beta')
(\bar {\partial} \beta' + cos(2\theta')\bar {\partial} \alpha')\cr-& 
{\sqrt{k'}\over{\pi}}e^{2(\Phi-\Phi_0)}(R_{3,2}-cos(2\theta_1)R_{2,3})
\partial x(\bar{\partial}\beta' + cos(2\theta')\bar
{\partial}\alpha')\Big\}}}
where $\Phi$ is the dilaton field defined in \hdilr.

Expressing the
matrix $R$ in terms of the  Euler coordinates as
\eqn\euler{ R = e^{\chi I_3}e^{\psi I_2}e^{\phi I_3}=
\left(\matrix{\cos\chi&\sin\chi&0\cr -\sin\chi&\cos\chi&0\cr 0&0&1\cr}\right)
\left(\matrix{\cos\psi&0&-\sin\psi\cr 0&1&0\cr\sin\psi&0&\cos\psi\cr}\right)
\left(\matrix{\cos\phi&\sin\phi&0\cr -\sin\phi&\cos\phi&0\cr 0&0&1\cr}\right)
}
where  $(I_i)_{j,k}=\epsilon_{ijk}$, we have for
the dilaton field~\foot{This is the same dilaton as in \gp\ 
(for a review, see \gpr).}
\eqn\hetdil{\eqalign{\Phi =& \Phi_0-{1\over 2} 
log\bigl[1+cos\psi cos(2\theta_1)cos(2\theta')
+(cos\psi cos\chi cos\phi-sin\chi sin\phi)cos(2\theta_1)\cr
&-(cos\psi sin\chi sin\phi- cos\chi cos\phi)cos(2\theta')\bigr]\cr
=&\Phi'_0-{1\over 2} log(cos^2\theta_1 sin^2 \theta'\cr
&+a^2cos^2\theta_1 cos^2\theta'
+b^2sin^2\theta_1 cos^2 \theta'
+c^2sin^2\theta_1 sin^2 \theta')}}
where $\Phi'_0=\Phi_0-{1\over 2}log\left[(1-cos\psi)(1-cos(\chi-\phi))\right]$
and
\eqn\aaa{a^2=
{(1+cos\psi)(1+cos(\chi+\phi))\over (1-cos\psi)(1-cos(\chi-\phi))}}
\eqn\bbb{b^2=
{1+cos(\chi-\phi)\over 1-cos(\chi-\phi)}}
\eqn\ccc{c^2=
{(1+cos\psi)(1-cos(\chi+\phi))\over (1-cos\psi)(1-cos(\chi-\phi))}}
The action in this parametrization is
\eqn\hetact{\eqalign{S&=\int d^2z\Big\{-{k\over{2\pi}}
\partial \theta_1 \bar {\partial} \theta_1 
+{k'\over{2\pi}}\left(\partial \theta' \bar {\partial}\theta'  
+\partial\alpha' \bar {\partial}\alpha' 
+ \partial \beta' \bar {\partial}\beta' 
+ 2cos(2\theta')\partial \alpha' \bar {\partial} \beta'\right)
+{1 \over {2 \pi}}\partial x \bar {\partial}x\cr
-&{k'\over{\pi}}\Big[{1 \over (1+cos \psi)[cos(2\theta_1)cos(2\theta')
+1+(cos(2\theta_1)+cos(2\theta'))cos(\chi+\phi)]}\cr&
\times \bigl[(1+cos \psi)
(cos(2\theta_1)+cos(\chi+\phi))\cr&-e^{2(\Phi-\Phi_0)}
sin^2 \psi(sin\phi-cos(2\theta_1)sin\chi)
(cos(2\theta_1)sin\phi-sin\chi)\bigr]\Big]\cr&\times (\partial \alpha' +
 cos(2\theta') \partial \beta')
(\bar {\partial} \beta' + cos(2\theta')\bar {\partial} \alpha')\cr-& 
{\sqrt{k'}\over{\pi}}e^{2(\Phi-\Phi_0)}
sin\psi(sin\phi-cos(2\theta_1)sin\chi)\partial x
(\bar{\partial}\beta' + cos(2\theta')\bar{\partial}\alpha')\Big\}}}
The coordinates $\alpha'$ and $\beta'$ as  $SU(2)$ Euler angles are compact.
$\alpha'+\beta'$ and $\alpha'-\beta'$ are defined modulo $2\pi$.
In regions for which $W>2$ replace $\theta_1$ by  $i\theta_{2}$.
For $W<-2$ replace ${\pi \over 2}-\theta_1$ by  $i\theta_{3}$. 

The model of ref. \nw\ is a special subset of the present family of models,
corresponding to $\psi = \pi$ ($a=c=0$). 
The parameter $\alpha$ there is related
to the present parametrization as $\alpha = \chi - \phi -{\pi \over 2}$
($b^2={1-\sin\alpha\over 1+\sin\alpha}$).

For large $k,k'$ and for small radius of the circle parametrized by $x$,
the action \hetact\ describes a $4$-dimensional space-time parametrized
by $(\theta_i,\theta',\alpha',\beta')$.
The  $5$-dimensional metric and antisymmetric tensor  read from \hetact, 
produce a corresponding $4$-dimensional structure via
the Kaluza-Klein mechanism.  
The term  proportional to 
$\partial x(\bar{\partial}\beta' + cos(2\theta')\bar {\partial}\alpha')$ 
gives rise in $4$ dimensions
to a $U(1)$ gauge field which couples to the momentum as well as to the
winding along the $x$ circle.
Equation \hetact\ describes only the patch
corresponding to $|W|\le 2$, the full model contains also the other 
patches which have the same metric with $\theta_1$ replaced by $i\theta_{2}$
or ${\pi \over 2}-i\theta_3$.

When $a=c=0$ the gauge field vanishes, and the $4$-dimensional background
takes the form (for $k=k'$ in regions with $|W|\le 2$):
\eqn\metrici{{1\over k}ds^2=-d\theta_1^2+d\theta'^2+{b^2\cot^2\theta'\over 
{b^2+\tan^2\theta_1 \cot^2\theta'}}d\lambda_-^2 
+ {\tan^2\theta_1 \over {b^2+ \tan^2\theta_1 \cot^2 \theta'}}d\lambda_+^2}
\eqn\bfi{B_{\lambda_+,\lambda_-}=
{kb^2 \over {b^2+\tan^2\theta_1 \cot^2\theta'}}}
\eqn\dili{\Phi=\Phi_0-{1\over 2}\log(\cos^2\theta_1 \sin^2 \theta' +
b^2\sin^2\theta_1 \cos^2\theta')}
where $\alpha' \pm  \beta'\equiv\lambda_{\pm}\in [0,2\pi)$ and,
again, in regions for which $|W|>2$ make the appropriate  replacement
for  $\theta_1$. 
This is the one parameter family of $4$-dimensional time-dependent 
backgrounds with two Abelian isometries discussed in \refs{\nw,\gp}.
It describes a series of closed, inhomogeneous expanding 
and recollapsing universes in $1+3$ dimensions.
At the time $\theta_1=0$ (modulo $\pi$) there is a big bang singularity,
while at $\theta_1={\pi\over 2}$ (modulo $\pi$) 
there is a big crunch singularity.

More precisely, at the time $\theta_1=0$ and on the surface
$\theta'=0$ there is a curvature singularity and the dilaton goes to infinity.
On the other hand, for generic $\theta'$ there is an orbifold singularity, and
the dilaton is finite. 
The orbifold singularity is a BTZ-like \refs{\btz,\bhtz},
namely, an orbifold of $SL(2,\IR)$ where we identify 
$g\simeq e^{\pi\sigma_3}ge^{-\pi\sigma_3}$, $g\in SL(2,\IR)$.~\foot{This can
be obtained from the discussion near eq. (3.5) in \egkr.}
Similarly, at the time $\theta_1=\pi/2$ there is a BTZ-like singularity of 
the axial type $g\simeq e^{\pi\sigma_3}ge^{\pi\sigma_3}$, 
except on the surface $\theta'=\pi/2$ where there is a curvature 
singularity and the dilaton blows up.

These universes are connected to non-compact static ``whiskers'' \egkr\
at the big bang/crunch singularities. The geometry of the whiskers
with $W>2$, attached at $\theta_1=0$, 
is described by the metric \metrici\ with 
$i\theta_2$ substituted for $\theta_1$. For any value of $\theta_1,
\lambda_+, \lambda_-$ this geometry is singular for 
$\theta'\in [0,{\pi \over 2}]$ satisfying
\eqn\dsing{cot \theta' ={b \over {tanh \theta_2}}}
The whisker is static and contains a singular ``domain wall,'' starting
at the big bang singularity at  $ \theta_2=0$ extending 
to infinity where $\theta_2$
tends to $\infty$. Crossing this wall in the whisker, the role of time 
is exchanged between $\lambda_+$ and $\lambda_-$. 
Similarly, the whiskers with $W<-2$ are described by the geometry 
\metrici\ with ${\pi \over 2}-i\theta_3$ substituted for $\theta_1$.
Again, there is a singular domain wall extending from the big crunch point
at $\theta_1={\pi \over 2}$ which is the same point as $\theta_3=0$,
towards infinity of $\theta_3$, which is described by the equation
\eqn\dsingh{cot \theta' ={b \over {coth \theta_3}}}

Non-zero parameters $a$ and $c$ \aaa, \ccc\  correspond to turning on two
components of an Abelian gauge field in the universe \metrici\ -- \dili.
Generically, these remove the big bang/crunch curvature
singularities \refs{\gp,\gmv}, as we shall discuss in section 3.
Such backgrounds can be described by a $3$-parameter sub-family
of $O(2,3,\IR)$ rotations of the direct product of a 
two-dimensional black-hole ($SL(2,\IR)/U(1)$) with a parafermion sigma-model 
($SU(2)/U(1)$) and a circle ($U(1)$)
\refs{\gp,\gmv} (for a review, see \gpr).

\subsec{Wavefunctions}

The realization of the model as  a quotient of a group manifold 
enables one to express vertex operators in terms of those of the 
ungauged WZW model \egkr. A typical vertex operator on $SL(2,\IR)
\times SU(2)\times U(1)$ which is unexcited in the $g,g'$
directions but may wind around the $x$ circle is of the form
\eqn\ver{V^{j,j'}_{m,m',q;\bar m , \bar m' ,\bar q} =
K^j _{m,\bar m}(g) D^{j'}_{m',\bar m'}(g')e^{i(qx_L+\bar q x_R)}}
Here $K^j _{m,\bar m}(g)$ is the matrix element of $g\in SL(2,\IR)$
in the  representation with the value 
$-j(j+1)$ for the Casimir operator, between states with eigenvalues $m$ and 
$\bar m$ for the infinitesimal generator corresponding to 
${1\over 2} \sigma_3$.~\foot{A  representation 
from the continuous series of the infinite cover of $SL(2,\IR)$ is not
completely determined by the value of $j$, an additional phase is 
required to specify it. Also for these representations to each 
value of $m$ corresponds a $2$-dimensional subspace; another $Z_2$
 variable is required to fix a state. In that case  $K^j _{m,\bar m}(g)$
means to depend also on those extra variables (see \egkr\ for details).} 
$D^{j'}_{m',\bar m'}(g')$ is similarly defined for $SU(2)$.

Applying  the gauge transformation \gatr\ to $(g,g',x)$ the operator 
$V^{j,j'}_{m,m',q;\bar m , \bar m' ,\bar q}$ gets multiplied by
\eqn\phase{exp[i({ m\over \sqrt{k}}\rho + { m'\over \sqrt{k'}}\rho'
+q\rho'' +{ \bar m\over \sqrt{k}}\tau + { \bar m'\over \sqrt{k'}}\tau'
+\bar q\tau'')]}
On the coset, only those vertex operators for which this phase equals
to $1$ are allowed. Taking \orth\ into account we get a 
 constraint on the allowed charges for a vertex operator. 
In matrix notation this reads
\eqn\conchar{({ m\over \sqrt{k}}, { m'\over \sqrt{k'}},q)R+
 ({\bar m\over \sqrt{k}}, {\bar m'\over \sqrt{k'}},\bar q)=0}
Here $m', \bar m'$ are quantized to be half integral and $q, \bar q$
are quantized on the Narain lattice $\Gamma^{1,1}$.

\newsec{Removing Singularities}

\subsec{Geometry}

As mentioned in section 2, when the gauge field vanishes ($a=c=0$)
the background \hetdil\ -- \hetact\ describes a $4$-dimensional closed 
universe \metrici\ -- \dili. This universe is singular at $\theta_1=\theta'=0$
(a big bang singularity) and at $\theta_1=\theta'={\pi\over 2}$
(a big crunch singularity). These singularities can be seen, for instance, 
by inspecting the behavior of the dilaton \hetdil\ when $a=c=0$.
Their origin is the presence of points on the group manifold
which are fixed under a continuous subgroup of the $U(1)^2$ gauge group
(see appendix A). Singular domain walls, 
extending into the whiskers, are attached
to these big bang singularities. 
The domain walls emerge where the $2$-dimensional 
gauged orbit becomes null \egkr\ (see appendix A). 
Turning on the gauge field generically removes the big bang/crunch
singularities and pushes the domain walls into the 
whiskers. Next we discuss such singularities by inspecting the
dilaton, and describe their relations with fixed points under 
subgroups of the gauged $U(1)^2$. 

The dilaton field \hetdil\ is generically regular throughout the region
$|W| \le 2$. It develops singularities at some special
points on the boundary of this region, only for some two-dimensional 
subsets of the three-dimensional parameter space $R$. At the point
$(\theta_1, \theta')=(0,0)$ there is a singularity in \hetdil\ for 
matrices $R$ with $\chi +\phi = \pi$ ($a=0$). At the point 
$(\theta_1, \theta')=({\pi \over 2},{\pi \over 2})$
there is a singularity for $\chi +\phi = 0$ ($c=0$). 
At the point $(\theta_1,\theta')=(0,{\pi\over 2})$ 
singularity appears when  $\chi -\phi = 0$ ($a,b,c\to\infty$).
Finally, at $(\theta_1, \theta') =({\pi \over 2},0)$ the subset 
of models with $\chi -\phi = \pi$ ($b=0$) form a singularity.

These singularities correspond to the presence of points on the original
group manifold which are fixed under some continuous subgroup of the 
$U(1)^2$ gauge group. Thus the point $(\theta_1, \theta')=(0,0)$ 
corresponds to $g(\theta)=g'(\theta')=1$. 
This point is invariant under the subgroup of the 
gauge transformations \gatr\ for which $\rho=-\tau, \rho'=-\tau'$
and $\rho''=0$. This is consistent with the condition \orth\ only 
for matrices $R$ for which there exists a vector of the form
$(\rho,\rho',0)$ which is rotated by $R$ into $(-\rho,-\rho',0)$. This implies
that this vector is the axis of rotation for the matrix $e^{\pi I_3}R$.
For a matrix of the form $e^{\chi I_3}e^{\psi I_2}e^{\phi I_3}$ 
the condition that the axis of rotation lies in the $(1,2)$ plane
is $\chi+\phi=0$. Hence a singularity at the point  $\theta_1 = \theta'=0$
develops only for $R$ with $\chi +\phi = \pi$. For the same reason
the point $(\theta_1 , \theta')=({\pi \over 2},{\pi \over 2})$,
corresponding to $g(\theta)=g'(\theta')=i\sigma_2$
which is fixed by transformations with  $\rho=\tau, \rho'=\tau'$
and $\rho''=0$, becomes singular only for models with  $\chi+\phi=0$.
Similarly, the point  $(\theta_1,\theta')=(0,{\pi\over 2})$ 
is invariant under a $U(1)$
subgroup of \gatr\ only for a matrix $R$ which takes the vector     
$(\rho,\rho',0)$ into $(-\rho,\rho',0)$. The matrix $e^{\pi I_2}R$
has then its axis in the $(1,2)$ plane. This implies  
the condition $\chi -\phi = 0$ on $R$. The same reasoning shows
that the condition for singularity at $(\theta_1,\theta') =({\pi \over 2},0)$
should be $\chi -\phi = \pi$. 

Yet smaller families of models correspond to the restrictions $\psi=0$
or $\psi=\pi$. At $\psi=0$ the parametrization \euler\ is redundant, 
there is no distinction between $\chi$ and $\phi$, rather we have a 
one-parameter family of angle $\chi + \phi$
 rotations in the  $(1,2)$ plane. For this family,
both the points  $(0,
{\pi\over 2})$ and  $({\pi \over 2},0)$ are
fixed points, hence singular, but not the points  $(0,0)$
and $({\pi \over 2},{\pi \over 2})$. The models corresponding to
$\psi=\pi$ ($a=c=0$) form another one-parameter family of angle  $\chi - \phi$
rotations followed by a reflection in the  $(1,2)$ plane. In this case,
 corresponding to ref. \nw, both the points  $(0,0)$ and 
 $({\pi \over 2},{\pi \over 2})$ are singular but not the points 
$(0,{\pi\over 2})$ and $({\pi \over 2},0)$.

In regions where $|W|>2$ the dilaton is given by \hdilr\ with 
$cos(2\theta_1)$ replaced either by $cosh(2\theta_{2})$ (when $W>2$) or
by $-cosh(2\theta_3)$ (when $W<-2$). For a given value of
$\theta'$ the dilaton becomes singular
when
\eqn\sing{cosh(2\theta_{2})=-{1+cos(2\theta')R_{2,2} \over{ 
R_{1,1}+cos(2\theta')R_{3,3}}}~\, \quad {\rm for} \,\,\, W>2~,}
\eqn\singtwo{cosh(2\theta_{3})={1+cos(2\theta')R_{2,2} \over{ 
R_{1,1}+cos(2\theta')R_{3,3}}}~\, \quad {\rm for} \,\,\, W<-2~.}
In particular, in  models for which
\eqn\nsingw{R_{1,1}>|R_{3,3}|}
there is  no singularity in the dilaton in the whiskers
with $W>2$ for any $\theta_2,\theta'$.
Similarly for $W<-2$ there will be no singularity if 
\eqn\nsingh{R_{1,1}<-|R_{3,3}|~.}
The condition \nsingw\ is equivalent to 
\eqn\abcsmw{a^2>b^2~, \qquad c^2<1~,} 
as  can be seen  from the relations  
\eqn\bbaa{{b^2\over a^2}=
{(1+R_{2,2})-(R_{1,1}+R_{3,3})\over (1+R_{2,2})+(R_{1,1}+R_{3,3})}}
\eqn\ccoo{c^2=
{(1-R_{2,2})-(R_{1,1}-R_{3,3})\over (1-R_{2,2})+(R_{1,1}-R_{3,3})}}
Similarly, condition \nsingh\ is the same as 
\eqn\abcsmh{c^2>1~, \qquad a^2<b^2~.}  
Hence, starting with a cosmological background \metrici\ -- \dili\
specified by the parameter $b$, we can gradually ``push,'' say, 
a big bang singularity and the domain wall singularity to which 
it is connected in the whisker~\foot{See \egkr\ for details.} with $W>2$
to infinity,
by turning on a gauge field, parametrized by $a,c$, keeping $|c|<1$
and increasing $|a|$ till it reaches $|a|=|b|$.
Once condition \abcsmw\ is obtained there is no singularity 
at all in the whisker. A domain wall singularity will still exist 
in the whiskers corresponding to $W<-2$.
The singularities are pushed to infinity in all the whiskers for the choice
\eqn\tnsing{R_{1,1}=R_{3,3}=0~,}
corresponding to $\psi={\pi \over 2}, \phi=0$ in \euler, or equivalently
\eqn\tabcns{a^2=b^2~,\qquad c^2=1~.}

A particularly simple background emerges when choosing out of the
models satisfying \tnsing\ the one with 
\eqn\cd{R=\left(\matrix{0&1&0\cr 0&0&1\cr 1&0&0\cr}\right)~.}
This corresponds to 
$\psi={\pi \over 2}, \phi=0,  \chi={\pi\over 2}$, or $a^2=b^2=c^2=1$.
For this background the dilaton becomes constant, $\Phi=\Phi_0$.
The action in the region where $|W|<2$ turns out to be
\eqn\pact{\eqalign{S&=\int d^2z\Big\{-{k\over{2\pi}}
\partial \theta_1 \bar {\partial} \theta_1 
+{k'\over{2\pi}}\left(\partial \theta' \bar {\partial}\theta'  
+\partial\alpha' \bar {\partial}\alpha' 
+ \partial \beta' \bar {\partial}\beta' 
+ 2cos(2\theta')\partial \alpha' \bar {\partial} \beta'\right)
+{1 \over {2 \pi}}\partial x \bar {\partial}x\cr 
+& {\sqrt{k'}\over{\pi}}cos(2\theta_1)
\partial x(\bar{\partial}\beta' + cos(2\theta')\bar
{\partial}\alpha')\Big\}}}
This corresponds to the ordinary metric and antisymmetric tensor
of $SU(2)$ times the one-dimensional time-like  $\theta_1$ direction 
times the $x$ circular fibre, with a gauge field. At the point 
$\theta_1=0$ the region $W>2$ is attached. There also the dilaton is 
constant and the action is \pact\ with $i\theta_2$ substituted for 
$\theta_1$. In this region $\theta_2$ is a space-like coordinate, 
while a combination of $x, \alpha$ and $\beta$ becomes  time-like. 
Similarly, another constant dilaton whisker parametrized by $\theta_3$
is attached at 
$\theta_1={\pi\over 2}$.

We have seen that for a generic $R$ matrix the points fixed by a 
continuous $U(1)$ subgroup of gauge transformations are removed,
together with their associated curvature and dilaton singularities. 
Yet the presence of a compact $x$ circle does not remove orbifold 
type of fixed points, namely, points fixed under a discrete, infinite 
subgroup of gauge transformations (see appendix A). 
Let $r$ be the radius of
the circle parametrized by $x$. The gauge parameters $\rho', \tau'$ in
\gatr\ are only defined modulo $2\pi/\sqrt{k'}$, while  $\rho'', \tau''$
are defined modulo $2\pi r$.   The $4$-dimensional  surface on the 
$SL(2)\times SU(2)\times U(1)$ group manifold corresponding to 
$g(\theta)=1$, i.e. $\theta =0$, is then invariant under a gauge transformation
of the form \gatr\ for which $\rho=-\tau, \rho' = 2\pi m /\sqrt{k'},
\tau' = 2\pi n /\sqrt{k'},\rho'' = 2\pi lr ,\tau'' = 2\pi sr$, for
$m,n,l,s$ integers. Such a gauge
transformation has also to satisfy \orth.
Our choice \hetv\ fixes $\tau''$ to $0$. The matrix $R$ has then to take
a vector of the form $(\tau, 2\pi n /\sqrt{k'}, 0)$ to the vector
$(-\tau, 2\pi m /\sqrt{k'}, 2\pi lr)$. If $r\sqrt{k'}$ and the elements
of $R$ are rational, then there exist large enough integers $m, n$
and $l$ which  satisfy this condition. Of course, if $m, n,l$
satisfy it, so do any multiplication of them by a common integer.
Hence the points on the surface corresponding
to $g(\theta)=1$ are fixed by an infinite, 
discrete set of gauge transformations. 
Otherwise, if the number $r\sqrt{k'}$ or some elements of $R$ 
are non-rational, any point on this 
surface is ``almost'' a fixed point in the sense that there exists a gauge 
transformation taking it to a point arbitrarily close to itself.
A similar surface of fixed, or almost fixed, points under a discrete subgroup
exists at $\theta=\pi/2$. However, eq. \hdilr\ implies that
these orbifold fixed points do not induce any singularity in the dilaton.   
      
The orbifold singularities are BTZ-like \refs{\btz,\bhtz}.
This can be seen, for instance, in the example \cd, where the curvature
singularities are removed completely, as follows.  
For simplicity, consider the case $k=k'$.
On the whisker, say $W>2$, the 4-dimensional slice at $\theta'=\pi/4$
has the line element:
\eqn\pacti{ds^2=k\left((d\alpha')^2
+d\theta_2^2+\cosh^2\theta_2d\lambda_-^2-\sinh^2\theta_2d\lambda_+^2
\right)~,}
where 
\eqn\lplm{\lambda_{\pm}=\beta'\pm {1\over\sqrt{k}}x~.}
This is obtained from eq. \pact\ with $\theta_1\to i\theta_2$, 
and at $\theta'=\pi/4$.
The background \pacti\ is the same as the ``whisker'' of an extended
BTZ black hole
(for instance, compare to eq. (37) in \mm) times an interval in $\alpha'$.
In particular, the singularity at $\theta_2=0$ is BTZ-like.

Finally, we note that when $x$ is non-compact, the orbifold singularities are
also removed. Equivalently, it is the compactification of the fifth 
direction $x$ which introduces the orbifold singularities.
When $x$ is non-compact such backgrounds have some similarities with the 
``null brane'' orbifolds introduced in \refs{\fs,\s}.

\subsec{Wavefunctions}

The wavefunctions $K^{j}_{m,\bar m}(g)$ develop a logarithmic
singularity~\foot{Recall
that in this case a certain linear combination \dvv\
of the wavefunctions is regular
and describes an incoming wave from the boundary of the whisker which
is fully reflected \egkr.} 
if and only if $|m|=|\bar m|$ (see \egkr\ for details).
In this subsection we show that singular uncharged wavefunctions exist
if and only if the closed universes have a big bang and/or 
big crunch curvature singularities.

If we require $m=\bar m$ for a vertex operator with $q=\bar q = 0$, 
eq. \conchar\  implies $m'=\pm \bar m'$. The $m'=\bar m'$ solution
is possible only for a matrix $R$ in \conchar\ such that $R^T$ takes 
the vector $(m,m',0)$ into $(-m,-m',0)$. This implies that the axis of 
 rotation of the matrix $e^{\pi I_3}R^T$ lies in the 
$(1,2)$ plane. For $R$ parametrized as in \euler\ this condition reads
$\chi+\phi=\pi$. As discussed in the previous subsection, the geometry 
corresponding to such an $R$ is singular at the point $(\theta_1,
\theta')=(0,0)$. Similarly the solution with
$m'=-\bar m'$ is possible for  a matrix $R$ with $\chi -\phi =0$,
which gives rise to a background geometry singular at $(\theta_1,
\theta')=(0,{\pi \over 2})$. The function  $K^j _{m,\bar m}(g)$
for $m=\bar m$ has indeed a logarithmic singularity at 
$\theta_1=0$ (see \refs{\dvv,\egkr} for details). 
We see then that, when $q=\bar q = 0$, condition \conchar\ 
allows for a singular behavior of the vertex operator only in the 
non-generic case of a singular background geometry. Under the same 
condition,  $q=\bar q = 0$, a vertex operator with $m=-\bar m$ 
must, by \conchar, have $m'=\pm \bar m'$. This is only possible
for an $R$ matrix with $\chi -\phi=\pi$ or $\chi+\phi=0$. Both 
cases give rise to a background geometry with singularity at
 $\theta_1={\pi \over 2}$.
Again, the vertex operator  $K^j _{m,\bar m}(g)$ with $m=-\bar m$
has a singularity at $\theta_1={\pi \over 2}$. 
The singular operator is allowed
only for a singular geometry.

If $q$ and $\bar q$ are non-zero, one can have $m =\pm \bar m$ together
with condition \conchar\ for many $R$ matrices which give rise to
background geometries without big bang/crunch curvature singularities. 
The vertex operator shows then a singular 
behavior at $\theta_1=0$ or $\theta_1={\pi \over 2}$ even when 
at these times there are no curvature singularities.
Note however that these  operators 
represent Kaluza-Klein ultra-heavy excitations from four-dimensional 
point of view.

\newsec{Summary}

In this paper we have turned on gauge fields in the 
four-dimensional family of extended universes \nw, \egkr\
(parametrized by $b^2={1-\sin\alpha\over 1+\sin\alpha}$ \refs{\nw,\gp,\gmv}). 
This was done within the exact CFT backgrounds corresponding to 
${SL(2)_k\times SU(2)_{k'}\times U(1)_x\over U(1)\times U(1)}$ quotients.
By a Kaluza-Klein reduction from five to four dimensions, one obtains a four
dimensional time-dependent background with an Abelian gauge field
(when $k,k'$ are much bigger than the compactification radius of $x$).

We found that turning on a generic gauge field
(parametrized by $(a,c)$) results in
pushing the curvature big bang/crunch singularities 
and the domain walls connected to them 
towards the boundary of the whiskers.
By tuning the gauge field ($|a|\to |b|$, $|c|\to 1$) 
the curvature singularities are removed completely 
and the dilaton is finite everywhere.

An orbifold singularity similar to an extended BTZ singularity remains at 
a time when a compact universe meets a whisker.
On the other hand, if $U(1)_x$ is non-compact, 
in which case the time-dependent background is five-dimensional, 
the orbifold singularities are removed.

Finally, using the methods of \egkr, we found that uncharged
incoming wavefunctions from the whiskers can be fully reflected 
if and only if there is a big bang/crunch curvature singularity,
from which they are scattered,  
where in particular the dilaton blows up.

\bigskip
\noindent{\bf Acknowledgements:}
We thank G. Veneziano for raising a question which led to this work.
We thank V. Balasubramanian, B. Kol, A. Konechny, D. Kutasov, E. Martinec, 
M. Ro\v cek, N. Seiberg and A. Sever for discussions. 
This work is supported in part by BSF -- American-Israel Bi-National 
Science Foundation, the Israel Academy of Sciences and Humanities --
Centers of Excellence Program, the German-Israel Bi-National Science 
Foundation, and the European RTN network HPRN-CT-2000-00122.
E.R. acknowledges also the support of the Miller Institute -- UC Berkeley. 

\appendix{A}{Singularities in Quotients}

In this appendix, we explain the relation between singularities 
in quotient CFT backgrounds 
and fixed points in the underlying manifold 
under the action of a subgroup of the gauged group.

The motion of a string on a sigma-model background is described by the 
the action 
\eqn\gac{\int d^2z E_{\mu,\nu} \partial x^{\mu} \bar {\partial} x^{\nu}~,}
where $x^{\mu}$ are some coordinates on the target-space manifold and 
\eqn\ee{E_{\mu,\nu}= G_{\mu,\nu}+B_{\mu,\nu}~.}
Here $ds^2=G_{\mu,\nu}dx^{\mu} dx^{\nu}$ is the line element
and $B_{\mu,\nu}dx^{\mu}\wedge dx^{\nu}$ a $2$-form. 
A dilaton $\Phi_0$ is also present.

Let $H$ be a $d$-dimensional isometry group of target-space. 
The action of $H$ is
generated by $d$ Killing vector fields $\xi_{(\alpha)}, \alpha=1,..,d$, 
which are generically independent.
Gauging away this action amounts to replacing the action \gac\ by
\eqn\gagac{\int d^2z E_{\mu,\nu} [\partial x^{\mu}+\sum_{\alpha}
A_{(\alpha)}\xi^{\mu}_{(\alpha)}] [\bar {\partial} x^{\nu}+\sum_{\beta}
\bar {A}_{(\beta)}\xi^{\nu}_{(\beta)}]~,}
where, as in section 2, $(A_{(\alpha)}, \bar {A}_{(\alpha)})$ are $d$ gauge
fields for the $d$ isometries.
Integrating out these gauge fields yields the effective $\tilde E=
\tilde G + \tilde B$ quadratic form corresponding to the gauged sigma-model
in the space of orbits in the underlying manifold 
generated by the action of $H$. 
Define the $d \times d$ matrix 
\eqn\mat{M_{(\alpha),(\beta)} = \xi^{\mu}_{(\alpha)}
E_{\mu,\nu}\xi^{\nu}_{(\beta)}~.}
The Gaussian integration of \gagac\ over the gauge fields gives:
\eqn\tee{ \tilde {E}_{\mu,\nu}= {E}_{\mu,\nu}-
\sum_{\alpha,\beta}{E}_{\mu,\rho}\xi^{\rho}_{(\alpha)}
(M^{-1})_{(\alpha),(\beta)}
\xi^{\tau}_{(\beta)}E_{\tau,\nu}~.}
By construction 
\eqn\loee{ \xi^{\mu}_{(\alpha)}\tilde E_{\mu,\nu}=
\tilde E_{\mu,\nu}\xi^{\nu}_{(\alpha)}=0}
for every $\alpha$, namely, the gauged action is insensitive to motion 
along the gauge orbits.
The contribution to the dilaton from this integration is
\eqn\tdil{\Phi=\Phi_0-{1\over 2} log (det M)~.}
At a point on the underlying manifold which is fixed under the action of
some continuous subgroup of $H$, the $d$ Killing vectors $\xi_{(\alpha)}$
are not independent. On the orbit corresponding to
such a point $detM=0$, hence both the dilaton $\Phi$
in \tdil\ and the quadratic form $\tilde E$ of \tee\ become singular.
 
If the original metric $G$ on the underlying manifold is 
not positive definite, a singularity may occur
even for points which are not fixed. At such a point
the $d$ Killing vectors are independent; no  combination 
of the $\xi_{(\alpha)}$ vanishes. Still, some combination 
$\xi=\sum_{\alpha} a_{(\alpha)}\xi_{(\alpha)}$ may become 
a non-zero null vector
which happens to satisfy $\xi^{\mu} E_{\mu,\nu}\xi^{\nu}_{(\alpha)}=0$
for every $\alpha$. Then again $det M=0$ and $\Phi$ and $\tilde {E}$
become singular. This is the origin of the ``domain wall'' singularities 
referred to in eqs. \dsing, \dsingh.

Unlike the behavior near points fixed under continuous symmetry or near
domain walls, where the dilaton blows up approaching them, points fixed
under a discrete subgroup have a different geometrical 
influence. At least as far as leading behavior in $\alpha'$ is concerned,
no sign of a singularity is felt arbitrarily close to the fixed point. The 
singular behavior occurs only at the fixed point itself. Such  are the 
orbifold type points discussed in section 3. 

\listrefs
\end